\documentclass[a4paper,11pt]{article}
\usepackage{pos}
\usepackage{graphicx}
\usepackage{subfigure}
\usepackage{amssymb}
\usepackage{pifont}
\usepackage{amsmath}
\RequirePackage{longtable}
\definecolor{blazeorange}{rgb}{1.0, 0.4, 0.0}

\title{Measuring the Cosmic X-ray Background accurately}

\author*{Hancheng Li}
\author{Nicolas Produit}
\author{Roland Walter}

\affiliation{University of Geneva, Department of Astronomy,\\
  16, Chemin d'Ecogia, 1290 Versoix, Switzerland}

\emailAdd{Hancheng.Li@unige.ch}

\abstract{
	 Measuring the Cosmic X-ray Background (CXB) is a key to understand the Active Galactic Nuclei population, their absorption distribution and their average spectra. However, hard X-ray instruments suffer from time-dependent backgrounds and cross-calibration issues. The uncertainty of the CXB normalization remain of the order of 20\%. To obtain a more accurate measurement, the Monitor Vsego Neba (MVN) instrument was built in Russia but not yet launched to the ISS. We follow the same ideas to develop a CXB detector made of four collimated spectrometers with a rotating obturator on top. The collimators block off-axis photons below 100 keV and the obturator modulates on-axis photons allowing to separate the CXB from the instrumental background. Our spectrometers are made of 20 mm thick CeBr$_{3}$ crystals on top of a SiPM array. One tube features a $\sim$20 cm$^2$ effective area and more energy coverage than MVN, leading to a CXB count rate improved by a factor of $\sim$10 and a statistical uncertainty $\sim$0.5\% on the CXB flux. A prototype is being built and we are seeking for a launch opportunity. 
}

\FullConference{37$^{\rm{th}}$ International Cosmic Ray Conference (ICRC 2021)\\
		July 12th -- 23rd, 2021\\
		Online -- Berlin, Germany}

\graphicspath{{./Figures/}}

\begin{document}
\maketitle

\section{Introduction}\label{sec:Introduction}

The Cosmic X-ray background (CXB) was first revealed during a rocket flight~\cite{Giacconi1962}. Its high isotropy, measured by \textit{Uhuru}, suggested an extragalactic origin~\cite{Giacconi2001}. Later, many soft X-ray missions (\textit{Chandra}, \textit{XMM-Newton}, \textit{Swift}, \textit{NuSTAR}, etc.) tried to resolve the CXB into discrete X-ray sources, and it turned out that Active Galactic Nuclei (AGNs) are the primary contributors to the CXB flux below 10 keV~\cite{Luo2017}. The CXB peak in the hard X-rays is lose to $\sim$30 keV, where measurements suffer from low sensitivity, rough spatial resolution and time-dependent backgrounds (cosmic ray activation, etc.) that are tricky to subtract. Some experiments used an on-board obturator~\cite{Gruber1999} or Earth occultations~\cite{Churazov2007} to separate the CXB from other components. However, the uncertainty on the CXB normalization remain an order of 20\%.

Synthesis models of the CXB rely on integrated AGN luminosity functions, obscuration distributions and spectral templates. The main source of uncertainty comes from the hard X-ray spectral templates, which are measured only for a small number of bright sources and that remains poorly constrained on average. Note that 95\% of the CXB is contributed by AGNs with redshifts smaller than 2. The fraction of highly obscured AGNs is still poorly known. Hard X-ray observations indicates that only 10-15\% of Seyfert 2s are Compton thick \cite{2016A&A...590A..49E,2019A&A...626A..40P} (note that the fraction of sources sources with N$_H>10^{26}$ cm$^{-2}$ is almost unconstrained). The uncertainties on the CXB hard X-ray spectral shape and of its normalization is a major source of difficulty for the synthesis models, and therefore for our knowledge of the accretion power in the Universe and of the importance of heavily obscured AGN in the Universe.

The CXB flux need therefore to be determined with a much better accuracy. In this context, the MVN (Monitor Vsego Neba) instrument was built by the Space Research Institute of the Russian Academy of Sciences~\cite{Serbinov2021} but not yet launched. We follow a similar idea for a collimated instrument surveying the sky through a rotating obturator. In section~\ref{sec:Construction} we introduce our initial instrument concept. Calibration ideas will be discussed in section~\ref{sec:Calibration}. In section~\ref{sec:Simulation}, we present instrument simulation based on the use of the Geant-4 framework~\cite{Geant42003}. Then we introduce the data analysis principle in section~\ref{subsec:Dynamic}, and finally provide some outlook in section~\ref{sec:Discussion}.

\section{Design}\label{sec:Construction}

\begin{figure}[!htbp]
	\begin{center}
		\includegraphics[width= 0.3\textwidth, height = 6cm ]{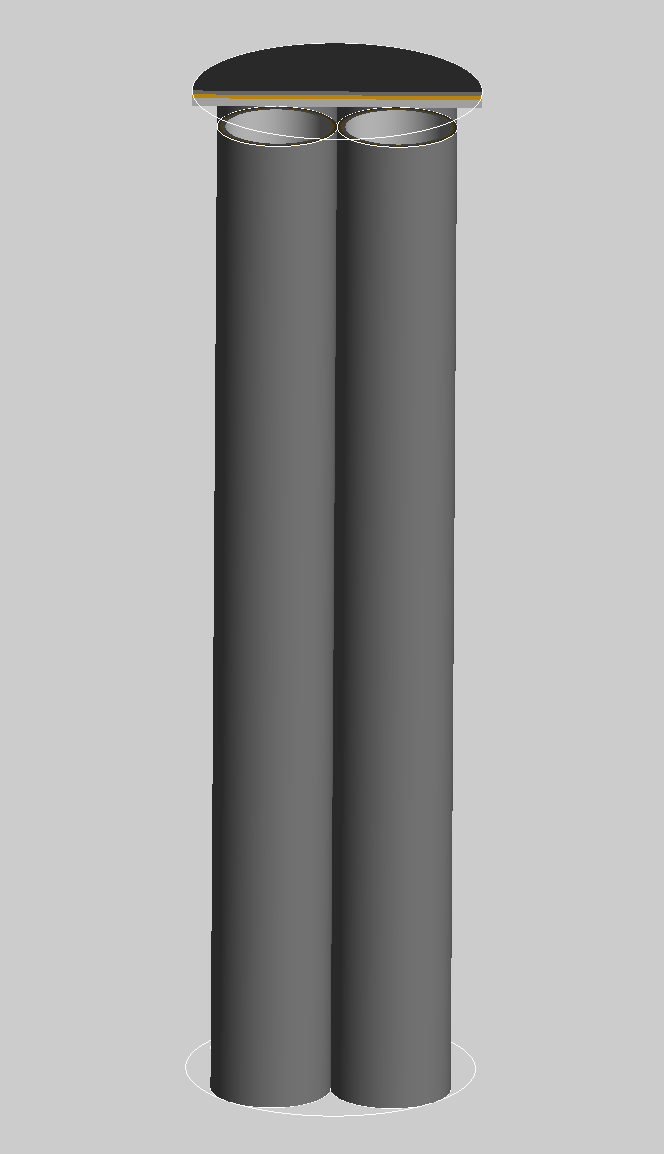}\includegraphics[width= 0.2\textwidth, height = 6.015cm ]{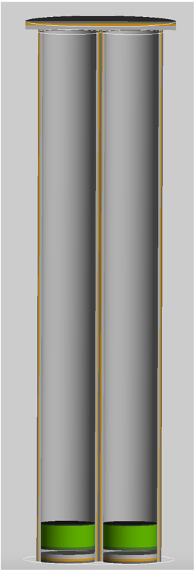}
		\caption{The left panel shows the obturator and four collimated spectrometers. The right panel is a vertical cut where the layered structure of the collimators, and the spectrometer detector modules could be seen. The length of one tube is 500 mm, and the inner diameter of the tube/spectrometer is 51 mm. The collimator sandwich is made by layers of 2 mm Al, 1 mm Cu and 1 mm Sn, from inside to outside respectively.}
		\label{fig:tube}
	\end{center}
\end{figure}

Our design is mainly consist of four collimated spectrometers with a rotating semicircular obturator on top of the aperture. The collimator tubes protect the spectrometers from receiving off axis photons before turning transparent at a certain energy threshold ($\sim$100 keV). The obturator will be periodically rotating, and shielding the aperture of the tubes, some of which will be close and the others open. This periodic change of the effective area of the spectrometers will modulate the count rates of the CXB and other emission coming from the zenith of the detector. All these components could therefore be modelled, form which the CXB flux could be extracted. A schematized design is shown in Figure~\ref{fig:tube}. The total weight is around 10 kilograms. The length (L) of one tube is 500 mm, and the inner diameter (D) of the tube/spectrometer is 51 mm. The Field of View (FoV) is around 26 square degrees (Full Width at Half Maximum).

\subsection{Collimator and obturator}\label{subsec:Collimator}

The collimator tubes are made of layers of Aluminium, Copper and Tin (from inner to outer). With thicknesses of 1-1-2 mm for the Sn-Cu-Al layers (off-axis projection is thicker by 1/$\sin \theta$), simulation gives $\sim$0\% penetration rate for photons with energy below $\sim$100 keV. 

The obturator is made of the same sandwich but twice thicker. The obturator and its contrarotation rotor minimizing the total torque will be driven by a simple DC motor. The position of the obturator will be read by an angular encoder. The rotation period will be later defined to adjust with the orbital period, so as to have a proper transit time of any sky region and to reach a correspondingly proper number of detected counts. The effective area of the collimator tubes versus the obturator opening angle are simulated in Figure~\ref{fig:arfmodulation}.

\subsection{\texorpdfstring{CeBr$_{3}$ crystal \& SiPM array detector}{lg}}\label{subsec:Crystal}

\begin{figure}[!htbp]
	\begin{center}
		\includegraphics[width= 0.6\textwidth, height = 6cm ]{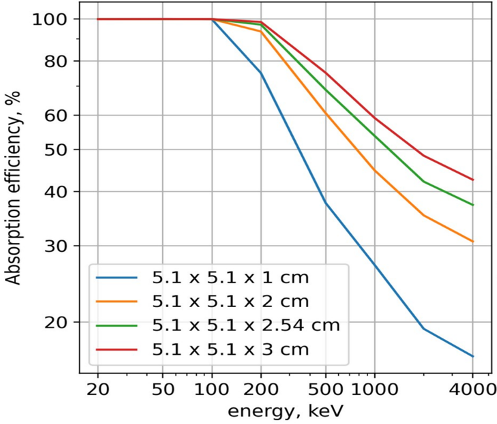}
		\caption{Absorption efficiency v.s. energy for different thicknesses of CeBr$_{3}$ (plot taken from~\cite{AnnaMPE2020}). X-axis is incident energy, Y-axis is absorption efficiency. Different colors are represented for different thicknesses. All curves start to drop down after 100 keV with different slopes. When the thickness of CeBr$_{3}$ is larger than 20 mm, its absorption efficiency remains $\sim$~100\% at 200 keV.}
		\label{fig:CeBr3}
	\end{center}
\end{figure}

The MVN uses 32 pixels (1 mm thickness) of CdTe crystal ~\cite{Serbinov2021} as sensitive detector with an excellent energy response at low energy, however it becomes transparent above $\sim$70 keV. We will use a detector made of CeBr$_{3}$ scintillation material and of a SiPM light detector. CeBr$_{3}$ provides improved detection performances~\cite{AnnaMPE2020}. Absorption efficiency as a function of energy for different thicknesses of CeBr$_{3}$ is shown in Figure~\ref{fig:CeBr3}. With a thickness of 20 mm of CeBr$_{3}$ our detector will keep an $\sim$100\% absorption efficiency up to 200 keV and still reaches a 60\% efficient at 511 keV (where we have emission line from a $\beta^+$ decay calibration source).

SiPM is increasingly used in space-borne detectors. It has large Photo Detection Efficiency (PDE), and its light yield lowers the low energy threshold. Other properties, such as good quantum efficiency, low bias voltage, compactness, robustness and insensitivity to magnetic fields, relax the detector design constraints. We will reuse the electronic developed for the 64 channels SiPM array of POLAR-2~\cite{Kole2021, NDA2021}. The size and position of each of the 64 channels should be further investigated. In order to reduce the impact of dark noise of SiPM and maintain a relatively low energy threshold, a cooling system is needed to reach ideally below -10 \textdegree C. 

\section{Calibration}\label{sec:Calibration}

Accurate knowledge about the detector's spectral properties, including Energy-Channel linearity, energy resolution, detection efficiency, etc., are crucial to measure the CXB normalization at the $\%$ level. We will therefore perform precise calibrations to the detector both on ground and in orbit.

\subsection{On ground}\label{subsec:ground}

Before launch, the detector will be calibrated by standard radioactive sources, such as $^{241}\rm Am$ and $^{22}\rm Na$, etc. The $\alpha$ decay of $^{241}\rm Am$ generates gamma rays peaking at 13.9, 17.8, 26.4 and 59.6 keV. And $^{22}\rm Na$ has $\beta^+$ decay emitting a positron immediately annihilating and releasing two gamma ray photons at 511 keV. By measuring the channel spectrum of these sources with the detector, a relationship between energy \& channel and energy resolution at different energy peaks will be obtained. The absolute detection efficiency at 511 keV can be measured by recording the single and coincident photon rate of the $^{22}\rm Na$ source. The absolute detection efficiency at other energies could then be cross calibrated using conventional sources. Geant4 simulations will verify the calibration procedures. Additionally, as the thermal conditions will change in orbit, we will measure the above values for different temperatures. 

\subsection{In orbit}\label{subsec:orbit}

There are two specific challenges in orbit. First the on ground calibrations will need to be checked as vibrations could change the detector geometry. Second the detectors will need to be monitored for aging and degradation due to radiation. We will use tagged $^{22}\rm Na$ and $^{241}\rm Am$ sources attached at the back of the obturator to periodically calibrate the energy calibration, resolution and the efficiency as it was done on ground. These calibrations, together with the Geant4 model that previously matched the on ground measurements, will monitor the aging of the detector. Additionally, we could also use the standard celestial object like the Crab to check the detection efficiency. As the high voltage (HV) is in a closed loop with the temperature, corrections of the detector responses versus HV and temperature could be applied. As the detector consisting of four identical tubes, the evolution of the cross calibration between the tubes will be checked.

\section{Instrument simulation}\label{sec:Simulation}

We used the Geant4 framework~\cite{Geant42003} to build the mass model of the detector presented in Section~\ref{sec:Construction}, to register the relevant physical mechanisms, and to inject incident photons from various directions and energies.

\begin{figure}[!htbp]
	\begin{center}
		\includegraphics[width= 1.0\textwidth, height = 6cm ]{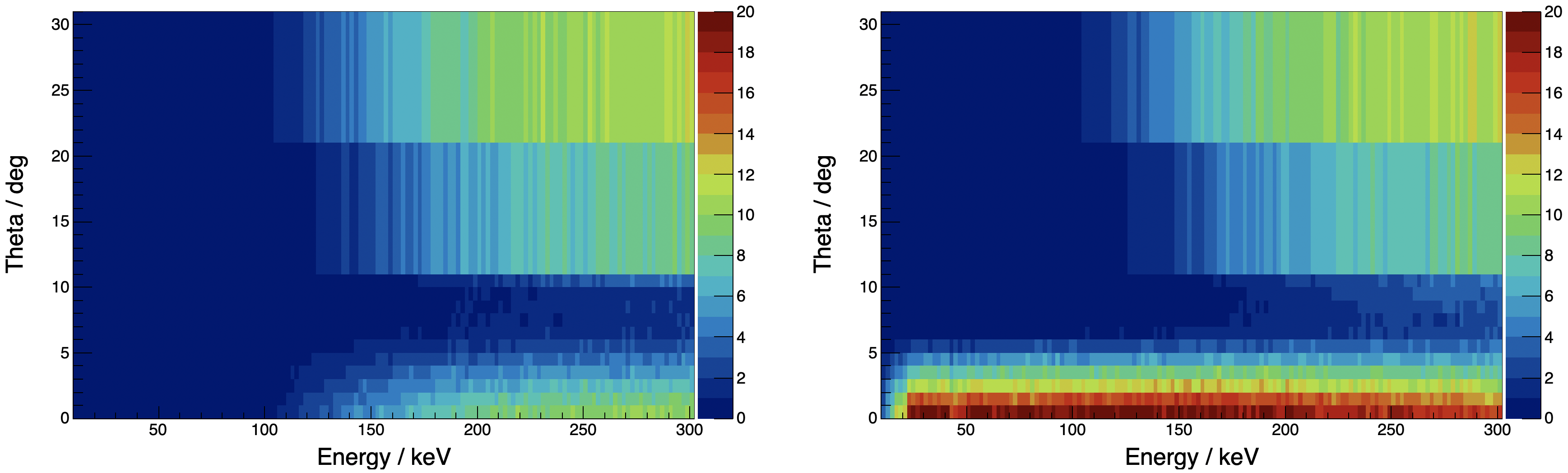}
		\caption{Effective area (color scale in cm$^2$) as a function of the off-axis angle theta and energy for one tube. The left panel is for an close tube. The right panel is for an open tube.}
		\label{fig:arf2d}
	\end{center}
\end{figure}

For the static simulations, the obturator is fixed and two tubes are fully close, and the other two open. Figure~\ref{fig:arf2d} shows the effective areas derived for the two types of tubes as a function of energy and off-axis angle up to 20\textdegree. The left plot indicates that the low energy threshold of the close tubes on-axis is $\sim$100 keV. The right plot indicates that the opening angle of the open tubes below 100 keV is $\sim$6 \textdegree. The energy response matrices of the detector are also obtained for different energies and incident angles.

\begin{figure}[!htbp]
	\begin{center}
		\includegraphics[width= 1.0\textwidth, height = 6cm ]{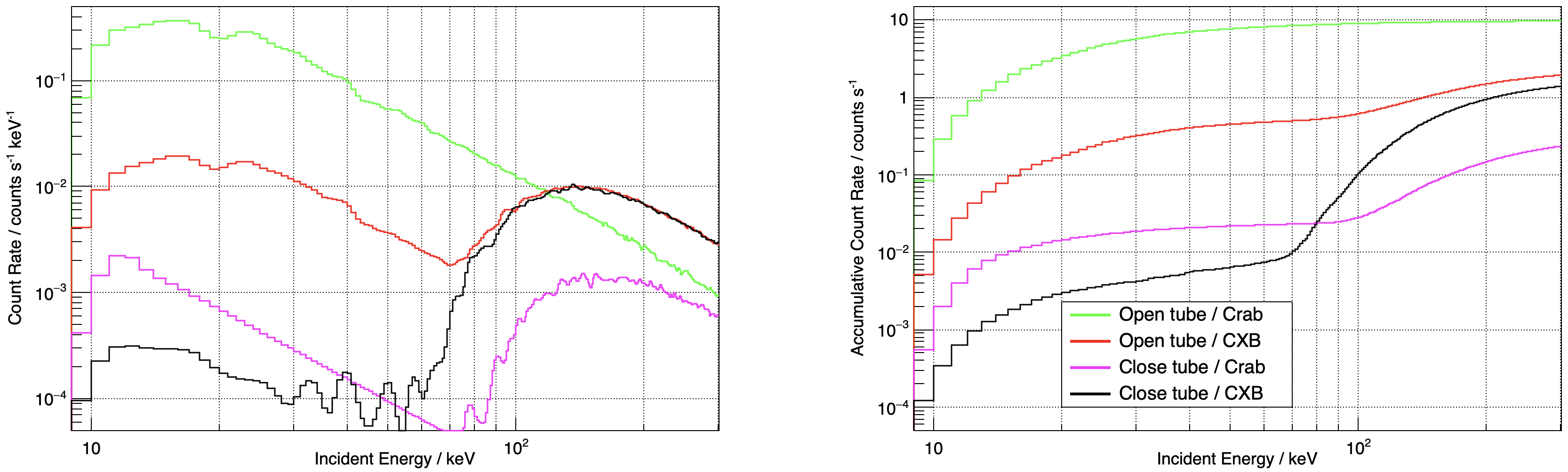}
		\caption{The left panel is the predicted count rate spectrum of the Crab and the CXB for the open and close tubes. The right panel shows the same count rate integrated over the energy.}
		\label{fig:countspec}
	\end{center}
\end{figure}

Using the simulated responses of the detector, together with the spectra of the CXB (from \cite{Gruber1999}) and of the Crab Nebula \& Pulsar (from \cite{Kuiper2001, Sizun2004}), we obtained the predicted count spectra shown in Figure~\ref{fig:countspec}.

The left panel of Figure~\ref{fig:countspec} shows the count rates of both the CXB and the Crab for a close tube. These sources will not faint to the close tube up to 100 keV. At higher energies, when photons starts to penetrate the obturator/collimator, the rate expected from the CXB becomes $\sim$10 times higher than that of the Crab. The count rates of the CXB and of the Crab are far higher in the open tube below 100 keV and about 10 times higher above that energy.

The right panel of Figure~\ref{fig:countspec} shows the count rate accumulated with energy, which is important to determine the statistical uncertainty achievable on the CXB normalisation. The total count rate of the CXB up to 70 keV is about $4.97\times10^{-1}$ counts/second in one open tube, while the CXB rate in the MVN instrument is $5.77\times10^{-2}$ counts/second). For the same orbital background as MVN and by using the Eq.(6) of ref.~\cite{Serbinov2021}, the statistical uncertainty of our CXB measurement will be of the order of $\sim$0.5\%.

\begin{figure}[!htbp]
	\begin{center}
		\includegraphics[width= 0.6\textwidth, height = 6cm ]{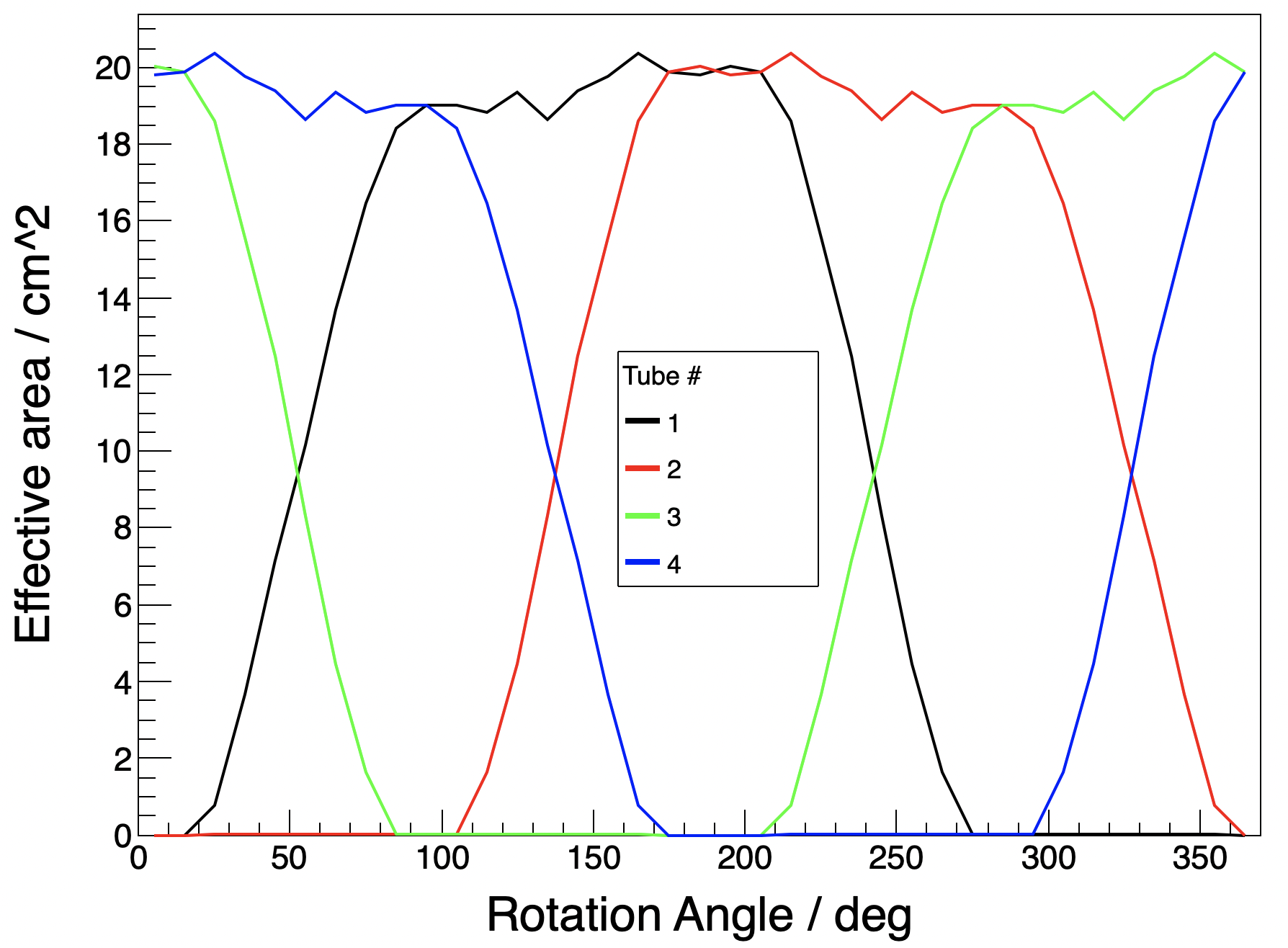}
		\caption{The effective area (per tube, at 30 keV) as a function of the rotation angle of the obturator.}
		\label{fig:arfmodulation}
	\end{center}
\end{figure}

Figure~\ref{fig:arfmodulation} shows how the effective area at 30 keV (around the peak of the CXB) is evolving with the rotation of the obturator. The effective areas of the four tubes periodically change leading to a modulation of the count rates, among which the maximum (fully open) and minimum (fully close) can be found from Figure~\ref{fig:countspec}.

\section{Data analysis principle}\label{subsec:Dynamic}

To reconstruct the fluxes of the CXB and of bright sources crossing the field of view regularly, we will need to accumulate data and to analyse all of them together taking into account all observational effects (boresight, obturator rotation phase), effective area, response and calibrations. 

The contribution of the internal instrumental background and of the on-board calibration sources can be subtracted easily using almost simultaneous measurements with close tube. The contamination of bright galactic celestial sources, could be measured and filtered out for the CXB measurements, or jointly modelling them with prior knowledge. Calibrations on the detector spectral properties and its aging in orbit, together with studies on the instrument background and how the data vary with observational parameters using three-year of data, will allow to reduce the systematic uncertainty of the CXB measurement to the percent level. 

The distribution of the CXB flux over the sky will also allow to measure the distribution of local AGN (the large scale structure of the local universe), and possibly detect the CXB dipole.

\section{Outlook}\label{sec:Discussion}

The instrument simulation shows that this detector design is able to separate the CXB from other emissions in 10-100 keV, and monitor the variability of bright celestial objects. The statistical uncertainty of CXB measurement will be of the order of $\sim$0.5\%. While the systematic uncertainty can be reduced into a few percent with 3 years observations. A comprehensive measurement simulation and data modeling algorithm will be performed in the next work to further study the systematic uncertainty. And a prototype is being built in our lab to perform various tests and verify that an absolute calibration could be obtained. A high accuracy calibration helps to reduce the systematic uncertainty. Meanwhile, we are seeking for a launch opportunity. If the space platform allows, several groups of tubes (four as one group) could be operated together, pointing at different directions simultaneously. The tubes \& obturator could also be thickened to reach a modulation of effective area at higher energy so as to extend our energy coverage for the CXB measurement. 


%
%
%


\begin{thebibliography}{99}

\bibitem{Giacconi1962} Giacconi, R. et al. 1962, \href{https://doi.org/10.1103/PhysRevLett.9.439}{\textit{Phys. Rev. Lett.}, 9, 439}

\bibitem{Giacconi2001} Giacconi, R. et al. 2001, \href{https://doi.org/10.1086/320222}{\textit{ApJ}, 551, 624}

\bibitem{Luo2017} Luo, B. et al. 2017, \href{https://doi.org/10.3847/1538-4365/228/1/2}{\textit{ApJS}, 228, 2}

\bibitem{Gruber1999} Gruber, D.~E. et al. 1999,  \href{https://doi.org/10.1086/307450}{\textit{ApJ}, 520, 
124}

\bibitem{Churazov2007} Churazov, E. et al. 2007, \href{https://doi.org/10.1051/0004-6361:20066230}{\textit{A\&A}, 467, 529-540}

\bibitem{2016A&A...590A..49E} Esposito, V. and Walter, R. 2016, \href{https://doi.org/10.1051/0004-6361:201527868} {\textit{A\&A}, 590, 49}

\bibitem{2019A&A...626A..40P} Panagiotou, C. and Walter, R. 2016, \href{https://doi.org/10.1051/0004-6361:201935052} {\textit{A\&A}, 621, 28}

\bibitem{Serbinov2021} Serbinov, D.~V. et al. 2021,  \href{https://doi.org/10.1007/s10686-021-09699-8}{\textit{Exp Astron}, 51, 493–514}

\bibitem{Geant42003} Agostinelli, S. et al. 2003,  \href{https://doi.org/10.1016/S0168-9002(03)01368-8}{\textit{NIM A}, 506, 250-303}

\bibitem{AnnaMPE2020} Ovisannikova A. 2020, Bachelor Thesis, TU Munich

\bibitem{Kole2021} Kole, M. 2021, 
\href{https://pos.sissa.it/395/600/}{\textit{PoS~(ICRC2021)}, 395, 600}

\bibitem{NDA2021} De Angelis, N. 2021, \href{https://pos.sissa.it/395/580/}{\textit{PoS~(ICRC2021)}, 395, 580}

\bibitem{Kuiper2001} Kuiper, L. et al. 2001,  \href{https://doi.org/10.1051/0004-6361:20011256}{\textit{A\&A}, 378, 918-935}

\bibitem{Sizun2004} Sizun, P. et al. 2004,  \href{https://ui.adsabs.harvard.edu/abs/2004ESASP.552..815S}{\textit{Proceedings of ESA SP-552}, 552, 815}

\end{thebibliography}
\end{document}